\documentclass[%
 reprint,
 amsmath,amssymb,
 aps,footinbib,showpacs,superscriptaddress,preprintnumbers,prl
]{revtex4-1}

\usepackage{graphicx}
\usepackage{dcolumn}
\usepackage{bm}

\begin{document}


\title{Origin of unexpected large Seebeck effect in SrTiO$_3$: nonperturbative polaron study from \textit{ab initio} cumulant expansion}

\author{Ji-Chang Ren}
\email{renj@njust.edu.cn}
\affiliation{Department of Physics, National University of Singapore, 117551 Singapore, Republic of Singapore}
\affiliation{Nano Structural Materials Center, School of Materials Science and Engineering, Nanjing University of Science and Technology, Nanjing 210094, Jiangsu, China}

\author{Jian-Sheng Wang}

\affiliation{Department of Physics, National University of Singapore, 117551 Singapore, Republic of Singapore}


%

\date{21 May 2018}

\begin{abstract}
The experimental unexpected large Seebeck coefficient in SrTiO$_3$ (STO) cannot be reproduced theoretically by the conventional Bloch-Boltzmann transport theory with electron-phonon coupling calculated perturbatively, indicating a failure of Fermi liquid picture in STO and ill-treatment of polaronic states. Starting from many-body interaction picture, the polaronic states can be precisely described by the method of cumulant expansion of retarded Green's function. By applying Kubo-Greenwood method, we found that the mysterious Seebeck coefficient can be fully described by a combination of polarons and Fermi liquid. The coexistence of two types of quasi-particles are attributed to the multi-band nature of t$_{2g}$ orbitals. Polarons are formed from the one with heavier electronic effective mass, while the one with lighter effective mass is renormalized into Fermi liquid quasi-particles. Our method provides a practical way to study the effect of strong electron-phonon coupling on transport properties from first principles. 
\end{abstract}

\maketitle


Electron-phonon coupling (EPC) in SrTiO$_3$ and related materials has profound impact on transport properties, such as superconductivity \cite{gor2016phonon,edge2015quantum,collignon2018metallicity,kim2011fermi} and hydrodynamics \cite{martelli2018thermal}. Another superior feature, ultra-high power factor \cite{scullin2008anomalously,jalan2010large,abutaha2015enhanced}, especially  large Seebeck coefficient \cite{ohta2007giant,king2014quasiparticle,okuda2001large}, has attracted increasing attentions for potential thermoelectric application at (or above) room temperature. Recently, based on experimental ARPES observation \cite{chang2010structure,wang2016tailoring,chen2015observation}, some dim satellite band structures have been found as a common feature for STO as well as related materials. Such satellites have been intuitively attributed to the polaronic states, which are originated from strong EPC and with their low lying energies just located in the transport regime, indicating their dominant role on the transport properties \cite{choi2015polaron}. However, the efforts of theoretical works are far behind experimental observation. One of the key reasons may be that the complex many-body interactions, especially strong EPC, are still unclear in STO.

For weak interacting particles, the transport properties can be described by Bloch-Boltzmann theory \cite{bloch1929quantenmechanik}, where the interactions are absorbed into scattering matrix \cite{sham1963electron}, or 
if only the diagonal part is considered, to momentum dependent parameters, the so-called single mode relaxation time approximation \cite{ziman1960electrons}. Within such approximations, calculated carrier mobility of STO is comparable with experimental observation \cite{himmetoglu2014first}. One would erroneously believe that the
transport properties can be safely described by Boltzmann single particle transport theory, even in strong coupling regime. However, the conclusion is not so straightforward. For interacting systems formed by Landau quasi-particles, Kadanoff \cite{prange1964transport} and Holstein \cite{holstein1964theory}, using Green{'s} function technique, proved decades ago that dc electric conductivity can be accurately calculated within the framework of Landau-Boltzmann theory after the cancellation of renormalization factors, but the effect of many-body interaction on Seebeck coefficient is debated through the past decades \cite{opsal1976electron,rathnayaka1985observation,lyo1977theory,jonson1990electron,chen2017large}. From a semi-classical point of view, by solving the Bloch-Boltzmann equation with relaxation time approximation, the Seebeck coefficient is $S=-k_B/e$$\int d\epsilon \Phi(\epsilon)\beta(\epsilon-\mu)(-{\partial f}/{\partial\epsilon})$$/$$\int d\epsilon \Phi(\epsilon)(-{\partial f}/{\partial\epsilon}) $$ $, where $f$ is the Fermi distribution function, $\beta=1/k_BT$, and a  transport function is defined by 
\begin{equation}
\Phi(\epsilon)=\frac{e^2}{V}\displaystyle\sum_{n\vec{k}} {\rm Tr}[v_{n\vec{k}}v_{n\vec{k}}]\tau_{n\vec{k}}\delta(\epsilon-\epsilon_{n\vec{k}}).
\end{equation} 
Here, $V$ is the volume of the system, $v_{\vec{k}}$ is group velocity of Bloch electron, $\epsilon_{n\vec{k}}$ is the eigenstate energy. 

In Eq.~(1), the interactions are totally absorbed into the relaxation time $\tau_{n\vec{k}}$. For small variation of $\tau_{n\vec{k}}$ in momentum space, the interactions can be appropriately cancelled out. Consequently, the value of Seebeck coefficient is dominated primarily by the non-interacting electronic structures. Based on this approximation, one can reach the conclusion that materials with the largest Seebeck should have extremely localized (delta-like) density of state (DOS) and moderate band gap \cite{mahan1996best}. However, this rule seems breakdown to explain the Seebeck effect in STO, which possesses an inferior band gap (3.25 eV) and conductive DOS far from delta-like function. Based on a three-band model, it is reported \cite{usui2010origin} that the Seebeck coefficient can be enhanced, but this individual effort is still too weak to compare with experimental values, indicating the failure of non-interacting picture for the description of Seebeck effect in STO.

Extensive experimental works \cite{chang2010structure,yamada2013measurement,devreese2010many,choi2010dimensional,cancellieri2016polaronic} have shown the existence of obvious EPC in intrinsic and slightly doped STO, forming the so-called Fr\"{o}hlich polarons. Its effect on transport properties, especially on Seebeck coefficient, is still illusive. It is known that the coupling strength of Fr\"{o}hlich polarons is at an intermediate regime. Can intermediate EPC be described by perturbation theory? Does quasi-particle picture still work in STO? What's the effect of Fr\"{o}hlich polarons on Seebeck? In this letter, we aim to answer these questions.

To study many-body electron-phonon interaction in STO, we discuss perturbation theory to start with. Firstly, due to forbidden momentum transfer, the deformation potential theory (DP) \cite{bardeen1950deformation} is not appropriate for STO. Then, we turn to Migdal perturbation approximation \cite{migdal1958interaction}, which contains the first order Fock term of Feynman diagrams. In many-body interaction picture, the Bloch eigenstates are replaced by spectral function with renormalized electronic energy and finite lifetime,
\begin{equation}
A_{n\vec{k}}(\omega)=-\frac{{2\rm Im}\Sigma(\omega)}{(\hbar \omega+\mu-\epsilon_{n\vec{k}}-{\rm Re}\Sigma(\omega))^2+{\rm Im}\Sigma(\omega)^2}.
\label{spectralfunction}
\end{equation}  
Here, the imaginary part of electron self-energy corresponds to lifetime of Landau quasi-particle. The real part is the renormalized energy corresponding to the central position of quasi-particle peak in spectral function. By setting the first part of denominator of Eq.~(2) to be zero, the renormalization energy can be obtained by self-consistently solving the equation. It should be noted that for the case of spectral functions with multi-peaks \cite{xu2013hidden}, i.e. multi-root of the equation, Landau quasi-particle picture may fail, as will be discussed later.   

By ignoring the transport vertex correction \cite{mahan2013many}, the transport relaxation time reduces to lifetime of quasi-particle which can be obtained by imaginary part of the electron self-energy. As shown in Fig.~1(c), at low energy range, the scattering rate increases rapidly. Such sensitive energy dependent relaxation time cannot be cancelled out appropriately for the calculation of the Seebeck coefficient. Further, in Fig.~1(d), it can be found that longitudinal optical (LO) modes contribute mostly to the scattering rate even at low energy range. Two step-like features appear at around 50 meV and 100 meV, which are the energy scales of two LO modes phonons. The electron-LO-phonon scattering is dominated by the absorption process when the electron energy is smaller than $\hbar\omega_{{\rm LO1}}$. Then, for $\epsilon_{n\vec{k}}>\omega_{{\rm LO1}}$, the LO-phonon emission process takes over the scattering as the electron energy increases. Due to the increase of phonon population, an upward shift of the scattering rate can be found at 500\,K. Since for transport properties the main contributions are taken from the low energy part, the scattering between heavier effective mass and LO phonons plays the dominant role on the Seebeck coefficient.
\begin{figure}[!t]
	\includegraphics[width=8.6cm,keepaspectratio]{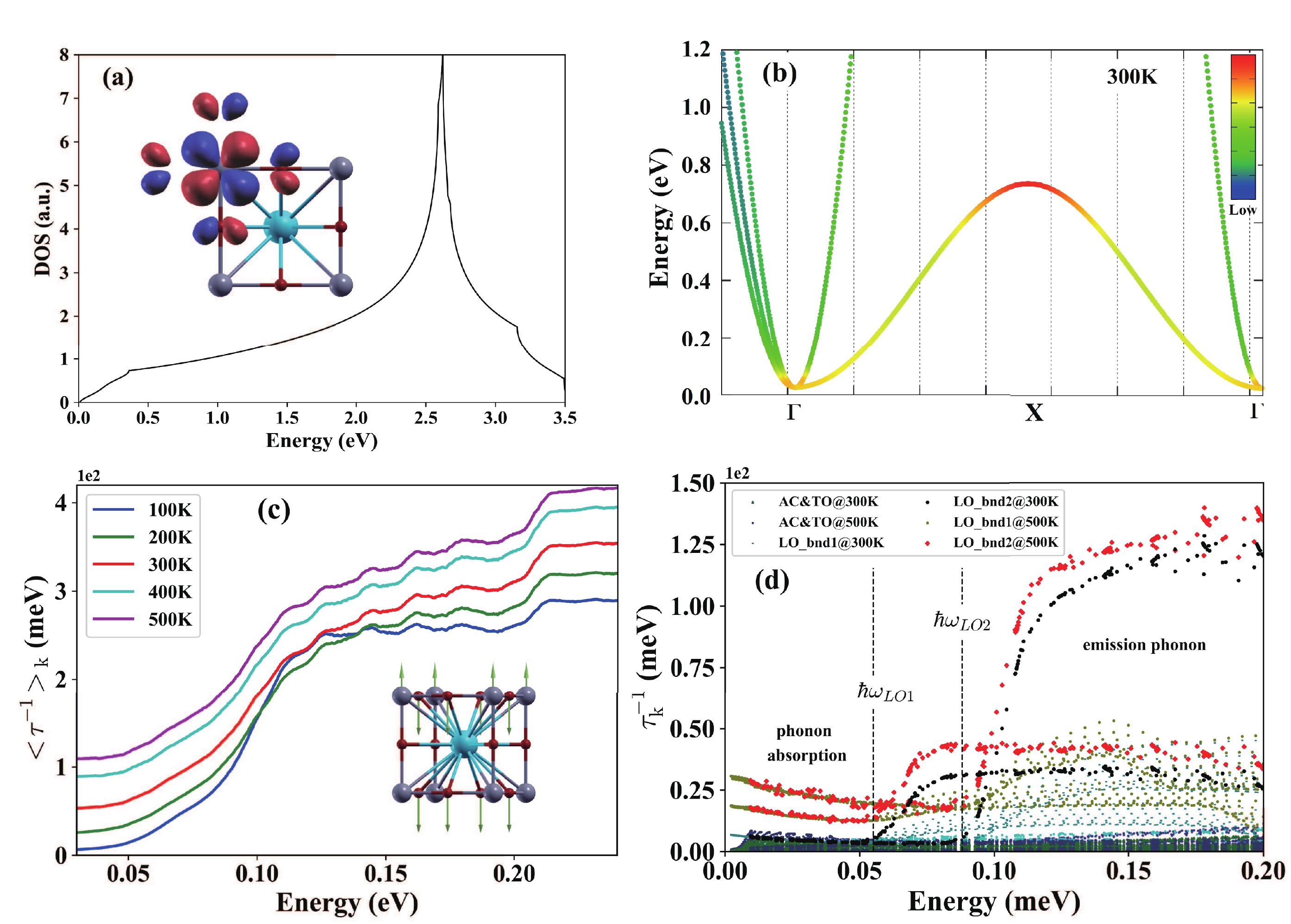}
	\caption{\label{fig:epsart} (a) DOS of t$_{2g}$ orbitals, insert is the wave function of t$_{2g}$ orbital. (b) Electron-acoustic phonon scattering rate distribution on band structure of t$_{2g}$ orbital. (c) Energy dependent (momentum averaged) relaxation time as a function of electronic energies under different temperatures. The insert shows the vibration of longitudinal phonon (LO) mode which contributes mostly to the scatterings. (d) Mode separated momentum dependent relaxation time at 300\,K and 500\,K. The two highest step-like features correspond to the LO modes with their energies denoted with dashed vertical lines, respectively.  
	}
\end{figure}
Unfortunately, within the framework of Bloch-Boltzmann theory, the obtained Seebeck coefficient drastically deviates from experimental results, as shown in Fig.~5. 
 
To reveal the many-body effect of the Seebeck coefficient, we go beyond the Migdal approximation to study the strong EPC in STO. Depending on the interaction strengths, the dressed carrier can be classified into small polaron and large polaron. Previously, we studied small polaron properties in EuTiO$_3$ \cite{chen2018strong}. However, for the case of STO, experimental evidence of optical conductivity \cite{van2008electron} suggests that the strength of EPC locates in the intermediate regime, originated from the coupling between electrons and LO modes of phonons (insert figure of Fig.~1(c)). The EPC in STO is described by  Fr\"{o}hlich Hamiltonian, which reads
\begin{equation}
\begin{aligned}
\hat{H} = \displaystyle\sum_{n\vec{k}}\epsilon_{n\vec{k}}\hat c^{\dag}_{n\vec{k}}c_{n\vec{k}} + \displaystyle\sum_{\lambda\vec{q}}\hbar\omega_{\lambda\vec{q}}\bigl(\hat a^{\dag}_{\lambda\vec{q}}a_{\lambda\vec{q}} + \frac{1}{2}\bigr) + \\ \frac{1}{\sqrt{N}}\sum_{\substack{n\vec{k},n'\vec{k'};\lambda\vec{q}}}g_{n\vec{k},n'\vec{k'}}^{\lambda\vec{q}}c^{\dag}_{n\vec{k}}c_{n'\vec{k}+\vec{q}}(a_{\lambda\vec{q}} + a^{\dag}_{\lambda\vec{-q}}),  
\end{aligned}
\end{equation}
where the third term is the coupling between electrons and phonons. Here $g_{n\vec{k},n'\vec{k'}}^{\lambda\vec{q}}$ is fully calculated from first principles with the QANTUM ESPRESSO code \cite{QE-2009}. Local density approximation (LDA) exchange-correlation functional of Perdew-Zunger type is chosen to describe electron-electron and electron-phonon interactions in the framework of density functional theory. Further, thanks to EPW code \cite{ponce2016epw}, Wannier interpolation techniques are applied to obtain EPC vertex $g_{n\vec{k},n'\vec{k'}}^{\lambda\vec{q}}$ in a fine $100\times100\times100$ k-mesh.

By taking the first order of diagrams into an exponential factor $C^{(1)}_{nk}$, the interaction part of Fr\"{o}hlich Hamiltonian contains exactly the first order Fock diagram and approximately part of other infinite number of diagrams. The final expression of $C^{(1)}_{nk}$ is obtained as:
\begin{figure}[!b]
	\includegraphics[width=8.6cm,keepaspectratio]{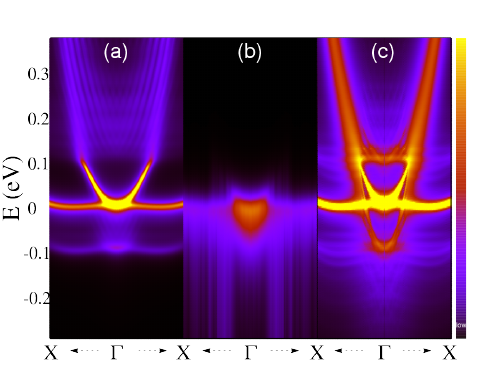}
	\caption{\label{fig:epsart} Momentum dependent spectral function at 300K with different methods: (a) Migdal approximation, (b) cumulant expansion time-ordered Green's function, (c) cumulant expansion retarded Green's function (RC). Fine features of polaronic states can be found at both side of Fermi level from cumulant expansion of retarded Green's function. The spectrum with dispersive-less feature around $\pm 0.1$ eV ($\pm 0.2$ eV) is polaronic states and the well-dispersive features are quasi-particle Fermi liquid states, indicating the coexistence of polaron and Fermi liquid states.}
\end{figure}
\begin{equation}
\begin{aligned}
C^{(1)}_{nk}(t)=\frac{-1}{N\hbar^2}\sum_{\substack{n'\vec{k'};\lambda\vec{q}}}|g_{n\vec{k},n'\vec{k'}}^{\lambda\vec{q}}|^2\Bigg[ (1+N_q-f_{n'\vec{k'}})\Big(\frac{it}{\Omega_{-}}+\\
\frac{1-e^{i\Omega_{-}t}}{\Omega_{-}^2}\Big)+(N_q+f_{n'\vec{k'}})\left(\frac{it}{\Omega_{+}}+\frac{1-e^{i\Omega_{+}t}}{\Omega_{+}^2}\right)\Bigg]
\label{CUMULANT}
\end{aligned}
\end{equation}
where $\Omega_{\mp}=(\epsilon_{n\vec{k}}-\epsilon_{n'\vec{k'}})/\hbar \mp \omega_{\lambda\vec{q}}$.
Then, the spectral function can be obtained
\begin{equation}
A_{nk}(\omega)=\frac{1}{\hbar}\int_{-\infty}^{+\infty}dt\, e^{-\frac{i}{\hbar}(\epsilon_{n\vec{k}}-\omega)t+C^{(1)}_{nk}(t)}.
\label{specofcum}
\end{equation} 

In Ref. \cite{verdi2017origin}, cumulant expansion method applied for time-ordered Green's function (TC) is calculated in energy domain by taking convolution to obtain the final spectral function, while our treatment is directly integrating the cumulant expanded retarded Green's function in time domain (RC), and then Fast Fourier Transform is taken to give the spectral function in energy domain. Technically, integration in time domain can naturally guarantee to fulfill normalization rule for the final spectral function. Theoretically, retarded Green's function is advantageous than the time-ordered one, as extensively discussed in recent theoretical works \cite{zhou2018cumulant,kas2014cumulant,mayers2016description,nery2018quasiparticles}. At zero temperature, the two methods become identical. 
\begin{figure}[!b]
	\includegraphics[width=8.6cm,keepaspectratio]{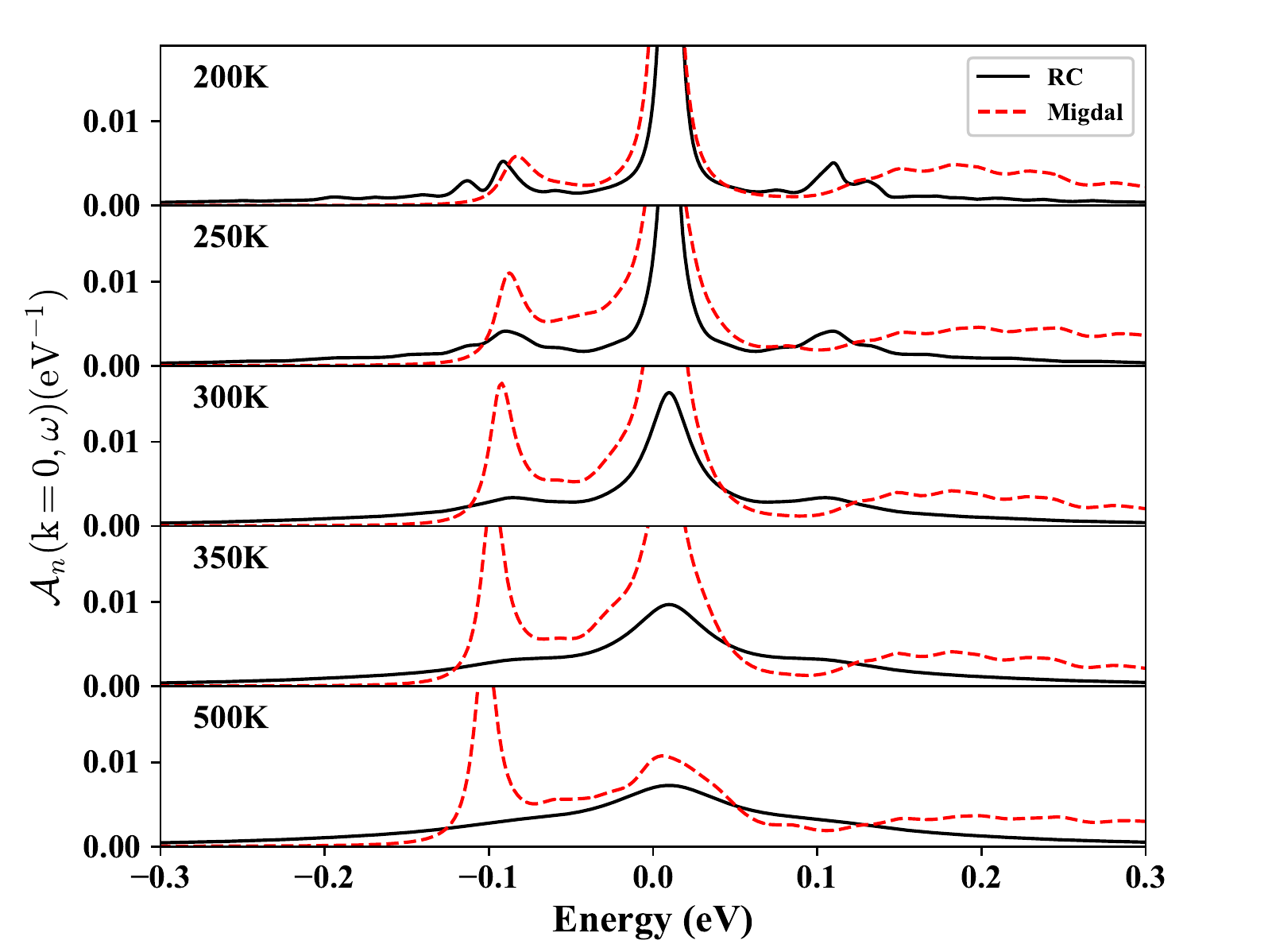}
	\caption{\label{fig:epsart} The behavior of single state spectral function calculated within Miagdal approximation (as implemented in EPW code) and RC method at $k=0$ with increasing of temperature, respectively. At 500\,K, in Migdal approximation, the peak of polaronic state is significantly larger than that of Fermi liquid state, indicating its failure at high temperature. While with RC method, the density of polaronic states is almost independent with the increasing temperature, and at high temperature, they are hidden behind the broaden Fermi liquid states.}
\end{figure}

Based on Eq.~(5), the spectral functions are calculated for STO at 300K. The fine features of the spectral function can be found by our RC method. As seen in Fig.~2(c), several satellites appear with a bright one at about ~0.1 eV and a dim one around ~0.2 eV, which are the identifications of polaronic features. Experimentally, APRES shows a satellite at around 0.17\,eV \cite{chang2010structure}, corresponding to the dim one in our theoretical calculations. While within Migdal approximation, only one bright satellite appears at 0.1\,eV, originated from the first order Fock diagram. Due to the perturbative nature of Migdal approximation, the high order non-perturbative polarons cannot be recovered in Fig.~2(a). On the other hand, by applying TC method, as implemented in the latest EPW code \cite{ponce2016epw}, Migdal approximation is obviously improved and the spectral function exhibits a slow decay behavior to the lower energy range, but the satellites features can hardly be found. Moreover, their TC method can only describe occupied states and the fine structures of the spectral function are totally lost. Since transport problem is very sensitive to the fine spectral function at lower energy range, TC method is not appropriate to describe the transport properties. 
\begin{figure}[!t]
	\includegraphics[width=8.6cm,keepaspectratio]{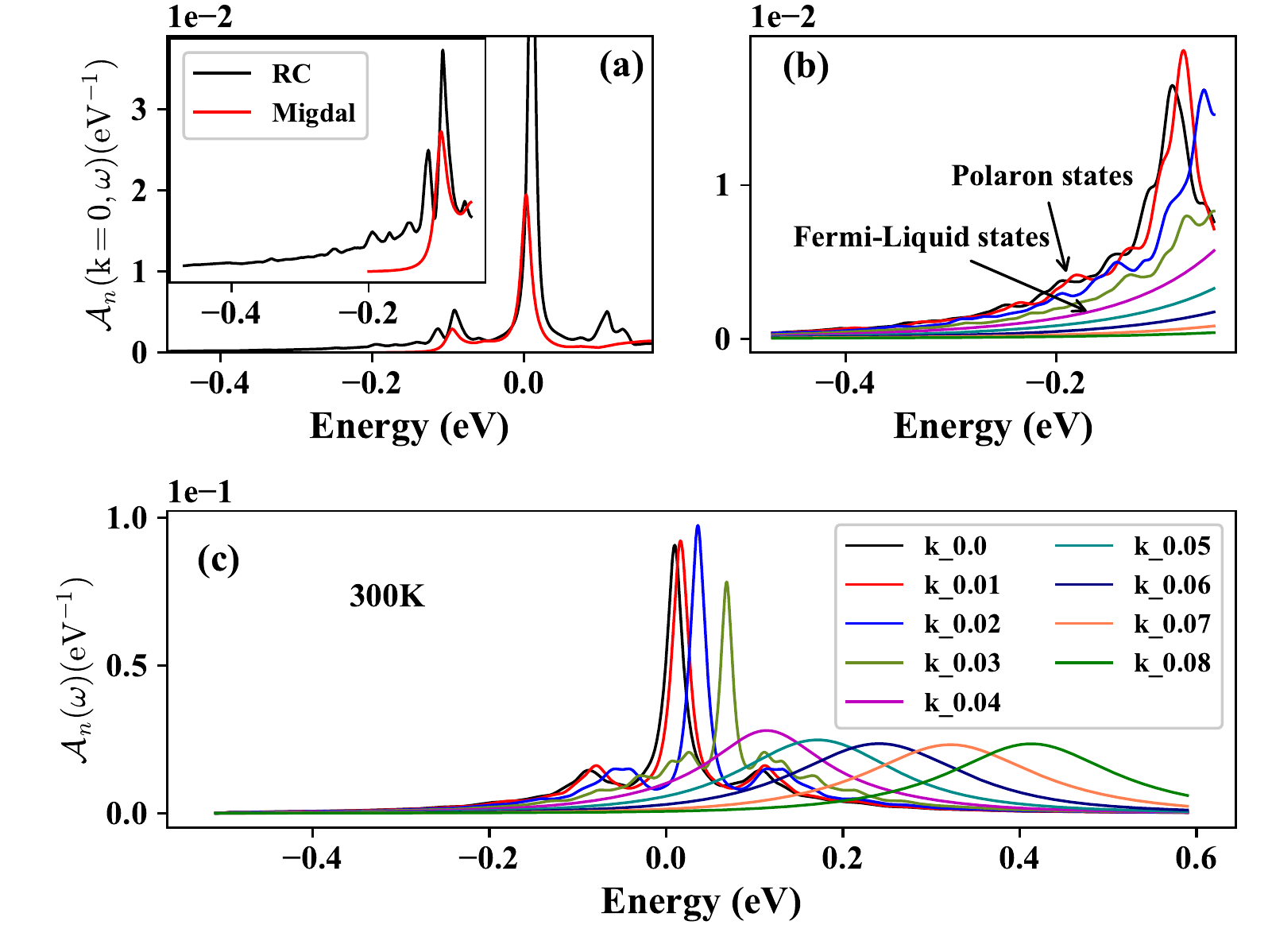}
	\caption{\label{fig:epsart} (a) Electronic spectral function with $k=0$ at 200 K within Migdal approximation and cumulant expansion method; (b) zoom in the tail of (c), both FLQP and polarons coexist in the transport energy range; (c) the spectral function with a list of k points, starting at $k=0$ and gradually moving away from BZ centre along X direction. A critical energy exists at 0.1\,eV, below which are polaron peaks while above ones are FLQP peak.}
\end{figure}

For a detailed investigation for the spectral function, we focus on a single state with $k=0$. As shown in Fig. 3, at low temperature, for both methods, a sharp quasi-particle peak and an obvious satellite are well separated and locate at energies slight above Fermi level and 0.1\,eV, respectively. The two peaks indicate distinct inherent physical nature. The first sharp peak corresponds to the quasi-particle mainly from Bloch state, so-called Fermi liquid quasi-particle (FLQP); the second lower peak is originated from polaronic states. The location of the satellite is just at the energy scale of highest LO mode of phonon, indicating that the polaron formation is originated from the coupling between electron and the highest LO mode of phonon. Due to the perturbative nature of Migdal approximation, only the largest polaronic states can be observed. With increasing temperature, the failure of Migdal approximation is even more obvious: the height of satellite increases rapidly with the rising temperature, and becomes significantly larger than FLQP peak, which is controversial with experimental observation \cite{chen2015observation}. By applying RC method, much more fine features can be obtained, as seen in Fig.4(a). The largest satellite repeats with an almost equal energy interval with 0.1\,eV, and companied by several relative small satellites nearby, forming by coupling with other phonon modes. These repeated multi-satellite features result in a significantly long tail of the spectral function, and such long tail may play a dominated role on the transport properties of STO.
\begin{figure}[!t]
	\includegraphics[width=8.6cm,keepaspectratio]{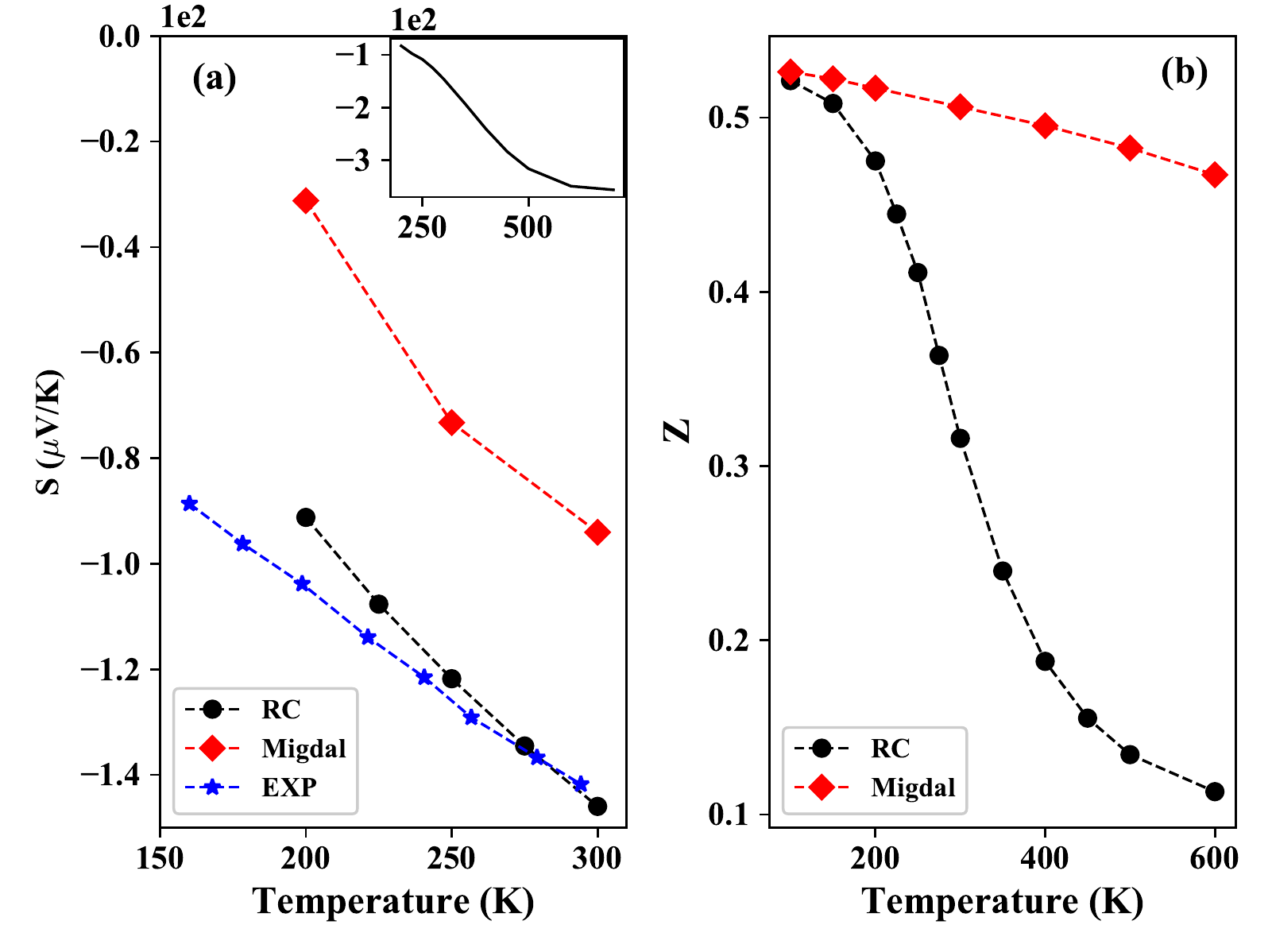}
	\caption{\label{fig:epsart} (a) Seebeck coefficient of STO calculated with Migdal approximation based on Landau-Boltzmann equation and RC method with Kubo-Greenwood formula, and compared with experimental result; insert shows the saturation appears at ~$700$ K. (b) renormalization factor, Z, defined as $ Z=[1-(\partial \rm Re\Sigma(\epsilon)/\partial \epsilon)]^{-1}|_{\epsilon=\epsilon_F}$, calculated with Migdal (Fork) interaction vertex and RC interaction vertex, respectively, as a function of temperature.}
\end{figure}

 With increasing temperatures, the behaviors of FLQP and polaron satellites exhibit different characteristics. The spectral functions of FLQP part broaden rapidly with increasing temperature, but the polaron satellites almost unchanged, and are gradually hidden under the broad distribution of FLQP when temperature beyond 350K. It is consistent with polaronic physics. Since the existence of polaronic states is not dependent on the population of phonons but the interaction strength between electrons and the self-induced polarization field (for Fr\"{o}hlich polaron), this coupling should be survived even at zero temperature. Recently, superconductivity has been observed experimentally for STO as well as related materials \cite{gor2016phonon,kim2011fermi}. Our results can provide an intuitive picture from many-body spectral function point of view: when decreasing temperature, FLQP will gradually shrink into delta-like function at some critical temperature, the long tail of the whole spectral function that dominates the transport properties is solely formed by polaronic states, for which new transport mechanism may come in and dominates the superconductivity properties in STO. 

Away from Brillouin zone center, the band degeneracy splits, as shown in Fig. 4(c), the EPC exhibits different effects on the band with light effective mass and the one with heavier effective mass. The spectral functions with single and more broaden peak is originated from the dispersive Bloch bands (light electrons). Others with satellite features are formed by the coupling between heavier electrons and LO modes of phonons. The general features of low energy range are formed by the ones with polaronic states. Interestingly, a critical energy that separates energy scales of FLQP and polarons states exists at 0.1\,eV, which is just the energy of the highest LO mode of phonon. This evidence further supports the FLQP nature at higher energy range ($>0.1$ eV). From semi-classic picture more LO phonon population will induces larger scattering rate for carriers, and results in the broaden of FLQP. However, such picture is breakdown for polaronic states ($<0.1$ eV). As a result, the tails of the spectral functions, as shown in Fig. 4(b), are largely contributed from polaronic states, but also contains obvious FLQP due to broadening of their spectral function. Thus, the transport nature of STO can neither primarily described by polaron picture nor from the framework of Landau-Boltzmann quasi-particle theory. In fact, two different quasi-particles both take part in the transport process. Thus, any types of renormalization that aims to map many-body interaction into quasi-particle picture may fail for the case of STO.
Since mapping multi-types of EPC into quasi-particle Landau-Boltzmann theory is problematic, we turn to deal with the Seebeck effect in STO fully in the framework of many-body transport theory. Starting from Kubo formula, by ignoring the transport vertex correction \cite{pruschke1993hubbard,palsson1998thermoelectric}, Kubo-Greenwood formula can be obtained: 
\begin{equation}
\begin{aligned}
T_{\alpha}= \int_{}^{}d\omega(-\frac{\partial f(\omega)}{\partial\omega}){\beta}^{\alpha}(\hbar\omega-\mu)^{\alpha}\Pi(\omega)
\label{CUMULANT}
\end{aligned}
\end{equation}
where $\mu$ is chemical potential, and $\Pi_{n}(\omega)$ is transport function, which reads
\begin{equation}
	\begin{aligned}
	\Pi(\omega)=\sum_{n\vec{k}} {\rm Tr}[v_{n\vec{k}}A_{n\vec{k}}(\omega)v_{n\vec{k}}A_{n\vec{k}}(\omega)]
	\end{aligned}
\end{equation}
Then, Seebeck coefficient can be obtained with $S=-(k_B/|e|)T_1/T_0$. With the Kubo-Greenwood formula, we calculated the Seebeck coefficient for slightly doped STO system. As shown in Fig.~5, by considering high orders of EPC diagrams, the Seebeck coefficient is significantly enhanced, compared with that from Migdal's perturbation theory, and in excellent agreement with experimental results\cite{okuda2001large}. Futher, since Kubo-Greenwood formula treats FLQP and polaronic states on equal footing, almost exact result can be obtained around room temperature. The slightly deviation at 200\,K may be attributed to the neglection of spin orbit interaction. Further, based on our calculations, we predict that the saturation temperature is as high as 700\,K and the highest Seebeck coefficient can be reached to $-356$ $\mu \rm {V/K}$ for STO with a doping level of $8\times10^{20}\rm {cm}^{-3}$. Our results indicate that the best thermoelectric performance of STO may be achieved for intermediate doping at high temperature regime.

To provide a quantitative estimation of many-body effect, we calculated the renormalization factor Z, as shown in Fig. 5(b). At 150\,K, the value of Z is about 0.55 within both Migdal and RC approximation, which is consistent with experimental value deduced from spectrum of optical absorption \cite{van2008electron}. With increasing temperature, the decrease of Z is unreasonably slow for Migdal approximation. After non-perturbative EPC comes in, the Z factor decreases with a nonlinear behavior when rising temperature: it decreases rapidly at low temperature range and approaches 0.3 at 300\,K, then decreases slowly towards to 0.1 at 600\,K, indicating quasi-particle is ill-defined at high temperature range.


In summary, based on first principles calculations, we have investigated the electron-phonon interactions in STO including higher orders of diagrams by applying cumulant expansion method. Multi-satellite features have been found for the spectral functions, corresponding to the polaronic states in STO. With Kubo-Greenwood formula, the Seebeck coefficient of STO was obtained by taking both the contributions from polarons and Fermi liquid quasi-particles. To the best of our knowledge, our calculations provided the most matching results with experimental data from a totally different point of view. In the end, we concluded that the Seebeck effect can be described neither purely by Fermi liquid nor primarily within polaronic picture but by mixing of the two. Since polarons are common feature in perovskite metal oxide materials, it appears that our method can be readily applied to study polaron effect on Seebeck coefficient of other perovskite oxides.

This work is supported by MOE tier 2 Grant R-144-000-349-112.

%

%

\end{document}